\documentclass[%
 reprint,
superscriptaddress,
 amsmath,amssymb,
 aps,
 prl,
]{revtex4-1}

\usepackage{graphicx}
\usepackage{dcolumn}
\usepackage{bm}
\usepackage{multirow}
\usepackage{threeparttable}
\usepackage{color}
\begin{document}

\title{Observation of the near-threshold intruder $0^-$ resonance in $^{12}$Be} 

\author{J.~Chen}
\affiliation{School of Physics and State Key Laboratory of Nuclear Physics and Technology, Peking University, Beijing 100871, China}

\affiliation{FRIB/NSCL Laboratory, Michigan State University, East Lansing, Michigan 48824, USA}

\author{S.~M.~Wang}
\affiliation{FRIB/NSCL Laboratory, Michigan State University, East Lansing, Michigan 48824, USA}

\author{H.~T.~Fortune}
\affiliation{Department of Physics and Astronomy, University of Pennsylvania, Philadelphia, Pennsylvania 19104, USA}

\author{J.~L.~Lou}
\email{jllou@pku.edu.cn}
\affiliation{%
 School of Physics and State Key Laboratory of Nuclear Physics and Technology, Peking University, Beijing 100871, China}%
 
\author{Y.~L.~Ye}
\affiliation{%
 School of Physics and State Key Laboratory of Nuclear Physics and Technology, Peking University, Beijing 100871, China}%

\author{Z.~H.~Li}
\affiliation{%
 School of Physics and State Key Laboratory of Nuclear Physics and Technology, Peking University, Beijing 100871, China}%
 
\author{N.~Michel}
\affiliation{%
 Institute of Modern Physics, Chinese Academy of Sciences, Lanzhou 730000, China}%
\affiliation{School of Nuclear Science and Technology, University of Chinese Academy of Sciences, Beijing 100049, China}

\author{J.~G.~Li}
\affiliation{%
 School of Physics and State Key Laboratory of Nuclear Physics and Technology, Peking University, Beijing 100871, China}%

\author{C.~X.~Yuan}
\affiliation{%
 Sino-French Institute of Nuclear Engineering and Technology, Sun Yat-Sen University, Zhuhai, 519082, Guangdong, China}%

 \author{Y.~C.~Ge}
\affiliation{%
 School of Physics and State Key Laboratory of Nuclear Physics and Technology, Peking University, Beijing 100871, China}%

\author{Q.~T.~Li}
\affiliation{%
 School of Physics and State Key Laboratory of Nuclear Physics and Technology, Peking University, Beijing 100871, China}%

\author{H.~Hua}
\affiliation{%
 School of Physics and State Key Laboratory of Nuclear Physics and Technology, Peking University, Beijing 100871, China}%

\author{D.~X.~Jiang}
\affiliation{%
 School of Physics and State Key Laboratory of Nuclear Physics and Technology, Peking University, Beijing 100871, China}%

\author{X.~F.~Yang}
\affiliation{%
 School of Physics and State Key Laboratory of Nuclear Physics and Technology, Peking University, Beijing 100871, China}%
 
 \author{D.~Y.~Pang}
\affiliation{%
 School of Physics and State Key Laboratory of Nuclear Physics and Technology, Peking University, Beijing 100871, China}%

\author{F.~R.~Xu}
\affiliation{%
 School of Physics and State Key Laboratory of Nuclear Physics and Technology, Peking University, Beijing 100871, China}%

\author{W.~Zuo}
\affiliation{%
 Institute of Modern Physics, Chinese Academy of Sciences, Lanzhou 730000, China}%
\affiliation{School of Nuclear Science and Technology, University of Chinese Academy of Sciences, Beijing 100049, China}

\author{J.~C.~Pei}
\affiliation{%
 School of Physics and State Key Laboratory of Nuclear Physics and Technology, Peking University, Beijing 100871, China}%

\author{J.~Li}
\affiliation{%
 School of Physics and State Key Laboratory of Nuclear Physics and Technology, Peking University, Beijing 100871, China}%

\author{W.~Jiang}
\affiliation{%
 School of Physics and State Key Laboratory of Nuclear Physics and Technology, Peking University, Beijing 100871, China}%
 
\author{Y.~L.~Sun}%
\affiliation{%
 School of Physics and State Key Laboratory of Nuclear Physics and Technology, Peking University, Beijing 100871, China}%

\author{H.~L.~Zang}%
\affiliation{%
 School of Physics and State Key Laboratory of Nuclear Physics and Technology, Peking University, Beijing 100871, China}%

\author{N.~Aoi}
\affiliation{%
 Research Center for Nuclear Physics, Osaka University, Osaka 567-0047, Japan}%
 
\author{H.~J.~Ong}

\affiliation{%
 Institute of Modern Physics, Chinese Academy of Sciences, Lanzhou 730000, China}%
 \affiliation{%
 Research Center for Nuclear Physics, Osaka University, Osaka 567-0047, Japan}%
 
\author{E.~Ideguchi}
\affiliation{%
 Research Center for Nuclear Physics, Osaka University, Osaka 567-0047, Japan}%

\author{Y.~Ayyad}
\affiliation{%
 Research Center for Nuclear Physics, Osaka University, Osaka 567-0047, Japan}%
\affiliation{FRIB/NSCL Laboratory, Michigan State University, East Lansing, Michigan 48824, USA}

 \author{K.~Hatanaka} 
\affiliation{%
 Research Center for Nuclear Physics, Osaka University, Osaka 567-0047, Japan}%

 \author{ D.~T.~Tran}
\affiliation{%
 Research Center for Nuclear Physics, Osaka University, Osaka 567-0047, Japan}%

\author{ D.~Bazin}
\affiliation{%
FRIB/NSCL Laboratory, Michigan State University, East Lansing, Michigan 48824, USA}

\author{J.~Lee}
\affiliation{RIKEN (Institute of Physical and Chemical Research), 2-1 Hirosawa, Wako, Saitama 351-0198, Japan}

\author{ Y.~N.~Zhang }
\affiliation{%
Department of Physics, Western Michigan University, Kalamazoo, MI 49008, USA}%

\author{J.~Wu}

\affiliation{%
 School of Physics and State Key Laboratory of Nuclear Physics and Technology, Peking University, Beijing 100871, China}%
 \affiliation{RIKEN (Institute of Physical and Chemical Research), 2-1 Hirosawa, Wako, Saitama 351-0198, Japan}

\author{H.~N.~Liu}

\affiliation{%
 School of Physics and State Key Laboratory of Nuclear Physics and Technology, Peking University, Beijing 100871, China}%
\affiliation{RIKEN (Institute of Physical and Chemical Research), 2-1 Hirosawa, Wako, Saitama 351-0198, Japan}

\author{C.~Wen}
\affiliation{%
 School of Physics and State Key Laboratory of Nuclear Physics and Technology, Peking University, Beijing 100871, China}%
 \affiliation{RIKEN (Institute of Physical and Chemical Research), 2-1 Hirosawa, Wako, Saitama 351-0198, Japan}

 \author{ T.~Yamamoto}
\affiliation{%
 Research Center for Nuclear Physics, Osaka University, Osaka 567-0047, Japan}%
 
  \author{ M.~Tanaka}
\affiliation{%
 Research Center for Nuclear Physics, Osaka University, Osaka 567-0047, Japan}%
 
  \author{ T.~Suzuki }
\affiliation{%
 Research Center for Nuclear Physics, Osaka University, Osaka 567-0047, Japan}%

\date{\today}

\begin{abstract}
A resonant state at  $3.21^{+0.12}_{-0.04}$\,MeV, located just above the one-neutron separation threshold, was observed for the first time in $^{12}$Be from the $^{11}$Be\,$(d,p)^{12}$Be one-neutron transfer reaction in inverse kinematics. 
 This state is assigned a spin-parity of  $0^-$, according to the distorted-wave Born approximation (DWBA) and decay-width analysis.  Gamow coupled-channel (GCC) and Gamow shell-model (GSM) calculations show the importance of the continuum-coupling, which dramatically influences the excitation energy and ordering of low-lying states. 
Various  exotic structures associated with  cross-shell intruding configurations  in $^{12}$Be and in its isotonic nucleus $^{11}$Li are comparably discussed. 

\end{abstract}

\maketitle

In weakly-bound nuclear systems approaching the drip-line, 
several exotic structures can emerge, including cross-shell intruder states and halo structures~\cite{Johnson2019}. Well known examples include the  ground states (g.s.) of the one-neutron halo nucleus $^{11}$Be~\cite{Talmi,Schmitt,Aumann} and the two-neutron halo nucleus $^{11}$Li~\cite{Tanihata,Simon1999}, which are characterized by the intrusion of the $2s_{1/2}$ orbital around the traditional neutron magic number $N=8$. 
This kind of orbital reordering should naturally lead to exotic structures not only in the g.s., but also in the low-lying exited states of the weak-binding nuclei. One outstanding example is $^{12}$Be, in which the low-lying excited states appear at 2.109\,($2_1^+$), 2.251\,($0_2^+$), and 2.715\,($1_1^-$)\,MeV~\cite{ Iwasaki2000,Shimoura2003,Imai2009, Alburger, Fortune1994}. These energies are dramatically reduced compared with  their counterparts in the isotone $^{14}$C~\cite{14C}, for which the lowest low-lying states are located at 6--7\,MeV, indicating strong intrusion of the $sd$-orbitals and the disappearance  of the $N=8$  magic  number.

In principle, for weakly-bound nuclei, low-lying intruder states may approach the particle-separation threshold or even appear as resonances above it. As a result, these single-particle states should strongly couple to the continuum, where  open quantum phenomena will emerge~\cite{Persson, Okolowicz2020,Mazumdar}. Although the reordering of some close-threshold orbits can be reproduced phenomenologically by the simple potential model (Woods-Saxon form)~\cite{Ozawa,Hoffman2016, Hoffman}, the systematic  description and understanding of the open quantum system require explicit treatment of the effective interactions and coupling to the continuum~\cite{Dobaczewski}.  Over the years, particular attention has been paid  to  narrow resonances at the vicinity of the particle-separation threshold~\cite{Okolowicz2020,Ikeda1968,YangZH,LiuY,JiangW}, associated with phenomena such as resonance trapping~\cite{Persson}, aligned eigenstates~\cite{Okolowicz2020} and Efimov states~\cite{Mazumdar}. 

 Traditional nuclear structure models~\cite{Yuan,WBP,Blanchon,Kanada,Garrido,Romero} 
have inherent difficulties to self-consistently account for the continuum coupling effects~\cite{Wang2019} and  to correctly reproduce the low-lying states in weakly-bound nuclear systems.  
In recent years, new theoretical approaches have been developed to incorporate the continuum-coupling effect, which greatly improves the description of the weakly-bound nuclei~\cite{Michel2009,Michel2002, Webb, Wang2019}. 
For instance, the Gamow coupled-channel\,(GCC) approach may handle  continuum-coupling  and  deformed core in a self-consistent way ~\cite{Webb, Wang2019}. 
The negative-parity states in weakly-bound nuclei around $N(Z) = 8$, which have cross-shell one-particle one-hole\,(1p-1h) configuration, should provide a sensitive testing ground to these theoretical calculations.

With one neutron in the intruder $2s_{1/2}$-orbital and the other in the normal $1p_{1/2}$-orbital, the $0^-$ and $1^-$ states in $^{12}$Be are expected to 
have extended spatial distributions and should strongly couple to the continuum~\cite{Johnson2019}. 
The $1^-$ state was observed at 2.715\,MeV~\cite{Iwasaki2000}, whereas  the $0^-$ state has been predicted by various theoretical approaches with an extremely large uncertainty, namely from 2.5 -- 9.0\,MeV~\cite{Yuan,WBP,Garrido,Romero,Blanchon,Kanada,Fortune201412Be}. 
However, in spite of a large number of experimental studies, no bound $0^-$ state below the one-neutron separation energy ($S_n = 3.17$\,MeV) has been found~\cite{Iwasaki2000,Shimoura2003,Imai2009, Alburger, Fortune1994,Johansen,Kanungo}. 
As outlined in~\cite{Fortune201412Be}, one neutron adding reaction on the nucleus $^{11}$Be, such as $^{11}$Be$(d, p)$, is a favorable  way to populate the 1p-1h unnatural-parity state in $^{12}$Be~\cite{Fortune201412Be,Garrido}.
However, the previous $^{11}$Be\,($d, p$) experiments at energies of 2.8~\cite{Johansen} and 5\,MeV/u~\cite{Kanungo}, focused on the bound states in $^{12}$Be and were insensitive to the coincident detection of  Be residues related to a $^{12}$Be resonance decay.

In this letter, we report on the first observation of the unbound $0^-$ state in $^{12}$Be from a  $^{11}$Be\,($d,p$) reaction at 26.9\,MeV/u using a specially
designed telescope centered 0$^{\circ}$. 
 The experiment  was carried out in inverse kinematics at the exotic nuclei (EN) beam line, Research Center for Nuclear Physics (RCNP), Osaka University~\cite{RCNP,RCNP1}. 
The secondary beam $^{11}$Be with an intensity of $10^4$ particle per second (pps) and a purity of $95\%$ was tracked onto a 4.00-mg/cm$^2$ (CD$_2)_n$ thick reaction target by using two upstream parallel-plate avalanche counters (PPACs).  A set of annular double-sided silicon detectors (ADSSD) was used to detect the protons from the $(d,p)$ reaction, covering the backward angular range of $135^{\circ}-165^{\circ}$ in the laboratory system.
Froward-moving Be isotopes (see inset of Fig.~1) were measured and identified by a downstream zero-degree telescope comprised of several layers of silicon detectors. 
The coincidence between the protons with the projectile-like Be ions emitting within $4^{\circ}$ is essential to select the reaction channel and to reduce the background. 
The protons were identified by their time of flight from the target to the annular silicon detector~\cite{CHEN2018,LiuY}. More details of the
experimental setup can be found in~\cite{CHEN2018}. 

The events identified as $^{12}$Be in the inset of Fig.~1 were used to select the $^{11}$Be\,$(d, p)$ reactions populating the bound states of $^{12}$Be as previously reported in ~\cite{CHEN2018}.  The $^{11}$Be and $^{10}$Be isotopes have much broader energy distributions, coming mostly from the reactions leading to one-neutron and two-neutron decays of the $^{12}$Be neutron-unbound states. 
Fig.~1 shows the excitation energy spectrum of $^{12}$Be, deduced from the energies and angles of the protons using the missing mass method and gated on the forward-moving $^{11}$Be or $^{10}$Be ions.  
The detection and calibration methods were validated from the study of the bound states as detailed in~\cite{CHEN2018}. 

\begin{figure}
  \includegraphics[width=1.0\columnwidth]{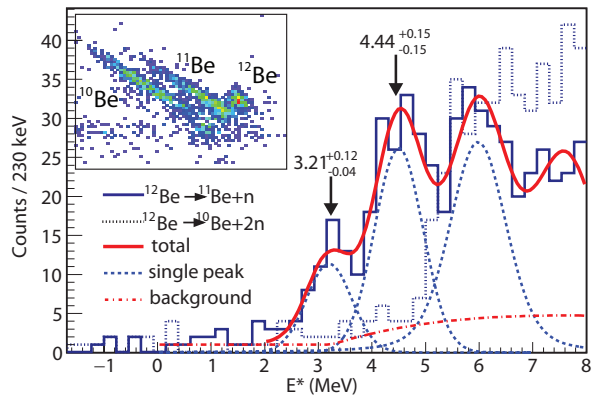}\\
  \caption{\label{fig:11BeEx} Missing mass spectra of $^{12}$Be unbound states, obtained from the protons in coincidence with $^{11}$Be\,(blue-solid spectrum) and $^{10}$Be\,(blue-dotted spectrum) ions detected at forward angles. The former is fitted with resonance peaks plus a simulated background.  
  The identified resonant  states are labelled with their corresponding excitation energies. 
  The inset shows in arbitrary units the energy loss versus remaining energy ($\Delta E-E$) spectrum of Be ions obtained from the zero-degree telescope 
  in  coincidence with the protons detected in the annular silicon array.  }
\end{figure}

\begin{table}
\caption{\label{tab:expt-S} Excitation energies, spin-parities and total decay widths for the low-lying resonances in $^{12}$Be observed in the present  $^{11}$Be\,$(d,p)^{12}$Be $\rightarrow$ $^{11}$Be$+n$ reaction. }
\begin{ruledtabular}
  \begin{threeparttable}
\begin{tabular}{ccccc}
\textrm{$E_x$\,(MeV))}&
\textrm{$\ell$}&
\textrm{$J^{\pi}$}&
\textrm{{$\Gamma$}\,(MeV)}\\
\colrule
   
   $3.21^{+0.12}_{-0.04} $ & 1 & $0^-$ & 0-0.37\\

   $4.44^{+0.15}_{-0.15}$ & 1 & ($2^-$) & 0-0.43\\

   \end{tabular}
   
    \end{threeparttable}
\end{ruledtabular}
\end{table}

The Breit-Wingner(BW) function~\cite{Lane,CAO2012} with energy dependent width is convoluted by the energy response function of the detection system to fit the excitation energy spectrum from the  $^{11}$Be\,$(d,p)^{12}$Be $\rightarrow ^{11}$Be$+n$ reaction. The energy response function at each excitation energy possesses   approximately a Gaussian shape with its width deduced from the simulation~\cite{CHEN2018}.
A resolution of 1.0\,--\,1.2\,MeV (FWHM), and a constant acceptance of $\sim$\,$86\%$ were obtained  for the states in the excitation energy range of $E^*=$2.0\,--\,6.0\,MeV.
The non-resonant background from direct breakup  process 
was simulated from random sampling of $^{11}$Be+$d\rightarrow\, ^{11}$Be+$n$+$p$ reaction in a three-body phase space and by tracking the particles into the detector system with smeared energies according to the energy resolution~\cite{Brown}. 
In addition, the random coincidence background was estimated using the event mixing method~\cite{YangZH}. 
 The excitation energy spectrum is fitted with the resonance energies and the corresponding widths left as free parameters by using the least-square method~\cite{Olive_2014}.
The  extracted results are listed in Table I and partially indicated in Fig.~1. 

The lowest energy peak at $3.21^{+0.12}_{-0.04}$\,MeV, is located just above $S_n$. The existence of this resonant peak was confirmed by applying various cuts on  $^{11}$Be isotopes (inset of Fig.~1) as well as on different angular ranges.  
The next peak at $4.44^{+0.15}_{-0.15}$\,MeV lies above two-neutron separation threshold, $S_{2n}=3.67$\,MeV. The relative branching ratios for its decay to one- and two-neutron final channels are determined by the ratios of the proton counts in coincidence with $^{11}$Be or $^{10}$Be ions, with the former being more than four times larger than the latter. 
Due to the complexity of the $^{10}$Be$+2n$ decay channel and the decreasing acceptance at higher excitation energies, we concentrate only on the analysis of the  $3.21^{+0.12}_{-0.04}$- and $4.44^{+0.15}_{-0.15}$-MeV peaks in the present work.

 The differential cross section shown in Fig.~2 was extracted from the corresponding resonant peak at each angular bin~\cite{LiuY}. 
The systematic uncertainty is estimated to be around $10\%$, contributed mainly from a reasonable variation of the $^{11}$Be cuts ($\sim$\,6$\%$), the influence of the fitting procedure ($\sim$\,$5\%$) and the uncertainty in target thickness ($\sim$\,$2\%$).  

The distorted-wave Born approximation (DWBA) calculations with transferred orbital angular momentum $\ell=1$ and $\ell=2$  (Fig.~2) were performed with an expanded version of the code DWUCK~\cite{Kunz}, modified by Comfort~\cite{Comfort} to include the Vincent-Fortune technique~\cite{Vincent} for stripping to unbound states. The curves for $\ell=0$  were calculated with a binding energy of 0.01 MeV~\cite{Vincent}. 
The optical model parameters (OMP) are the same as in~\cite{CHEN2018}.  
 For the $3.21^{+0.12}_{-0.04}$-MeV state, the determination between $\ell=0$ or $1$ seems marginal, with the corresponding minimum $\chi^2$ values of 1.54 and 0.23, respectively.  However,  the $\ell=0$ transfer to the $3.21^{+0.12}_{-0.04}$-MeV state can be excluded by the systematics of the level scheme in the $N =8$ isotones (refer to Fig.~5) and by the extracted narrow resonance width~\cite{Bertsch,Simon2007,swavechi}. 
 An $\ell=1$ assignment limits the possible spin-parities to $0^-$, $1^-$ or $2^-$.  The $1^-$ state has already been experimentally identified  at 2.71\,MeV~\cite{Iwasaki2000}, whereas the $2^-$ state should lie at much higher energy than the $0^-$ state, similar to the case of $^{14}$C~\cite{Fortune201412Be} and also considering the theoretical calculations shown in Fig.~3.  
 Hence, the $3.21^{+0.12}_{-0.04}$-MeV state is assigned a spin-parity of $0^-$. Based on comparison between the measured differential cross sections and the DWBA calculation~\cite{CHEN2018,ChenPRC},  a spectroscopic factor $(S)$ of 0.78(23) is deduced. This state possesses predominantly a cross-shell configuration $^{11}$Be$(1/2^+) \otimes (1p_{1/2})$, similar to the $1^-$ state, consistent with its large $S$ value in the $^{11}$Be\,$(d,p)$ reaction. 
Combined with the following theoretical discussions, the $4.44^{+0.15}_{-0.15}$-MeV state is tentatively assigned a single $2^-$ state with $S=0.25(5)$.  
The uncertainty of $S$ is typically around 20$\%$, resulted mainly from the variation of the OMPs~\cite{Kay2013}. 

\begin{figure}
  \includegraphics[width=1.0\columnwidth]{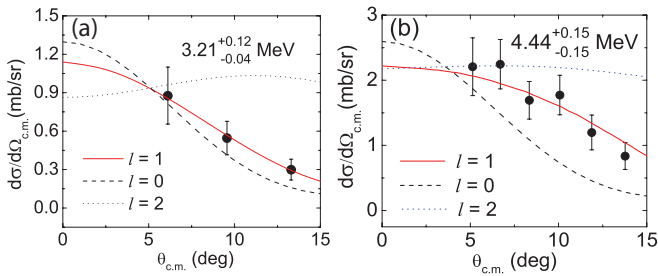}\\
  \caption{Differential cross sections of the $^{11}$Be\,$(d,p)^{12}$Be reaction populating  the (a) $3.21^{+0.12}_{-0.04}$-MeV and (b) $4.44^{+0.15}_{-0.15}$-MeV states of $^{12}$Be and the corresponding DWBA calculations. The error bars are statistical only.} \label{protonE}
\end{figure}

\begin{figure}
  \includegraphics[width=1.0\columnwidth]{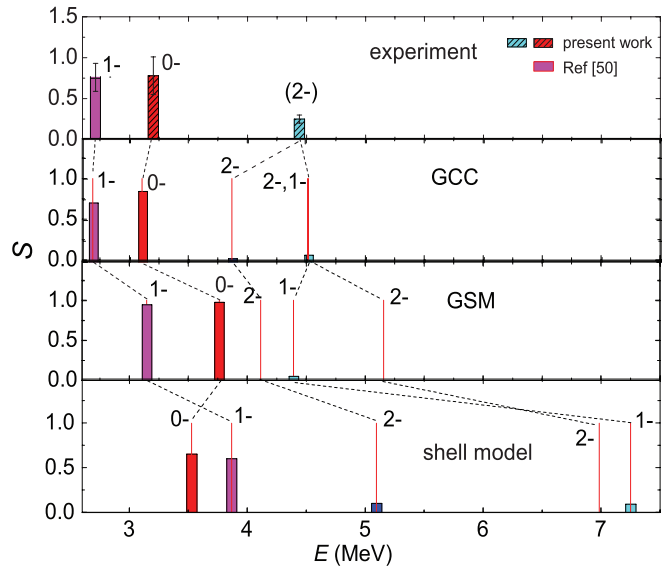}\\
  \caption{Excitation energies (located at the thin red solid vertical  lines) of the negative-parity states  and the corresponding $S$ values (height of the red bars) calculated by the GCC, GSM and shell model approaches in comparison with the present and previous experimental result. In addition, these three models predict the $0_3^+$ state at 4.8, 4.7 and 5.6 MeV (not plotted in the figure), respectively.
  }\label{protonE}
\end{figure}

In Fig.~3, the calculated energy spectra and $S$ values are compared with the experimental results. The latest YSOX interaction~\cite{Yuan,CHEN2018} was used in the shell model calculations. 
The energies of the  $1^-$ and $0^-$ states are predicted much higher than the experimental value and  are inverted with respect to each other. 
As a matter of fact, considering the weakly-bound and intruding mechanisms of $^{12}$Be,  its low-lying negative-parity states should be dominated by ($s$, $p$) or ($p$, $d$) configurations and thus strongly coupled to the continuum. This coupling is difficult to be dealt with using the shell model based on the harmonic oscillator (HO) basis. 
Gamow shell-model (GSM) incorporates continuum-coupling  by assuming a $^{8}$He core mimicked by a Woods-Saxon (WS) potential with an optimized  Furutani-Horiuchi-Tamagaki (FHT)  force~\cite{Furutani}. 
As shown in Fig.~3, the excitation energies of the $1^-$ and $2^-$ states are clearly reduced and the level ordering of the $1^-$ and $0^-$ states is restored,  showing the importance of  the continuum-coupling. 
However, there still exist some considerable discrepancies between the GSM calculations and the observations, possibly originating from the insufficiency of realistic interaction.  

For the negative-parity  $0^-$ and $1^-$ states populated with a significant $\ell=1$ strength in the one-neutron adding reaction on the $s$-wave intruding $^{11}$Be g.s., it is natural to expect a dominant three-body structure with two weakly-bound neutrons occupying  $1p_{1/2}$ and $2s_{1/2}$ orbitals coupled to the $^{10}$Be core. Accordingly, we performed calculations within Gamow coupled-channel (GCC) approach assuming a system of $^{10}$Be core plus two neutrons, together with  the continuum-coupling~\cite{Wang2019,Webb}. 
The calculation uses the same interaction and model space as in~\cite{Wang2019}. We choose $^{10}$Be as a deformed core including a non-adiabatic coupling with low-lying rotational states. The effective core-valence potential and the interaction between the two valence nucleons are taken in a deformed WS shape including the spherical spin-orbit term and the finite-range Minnesota force, respectively. 
This interaction has successfully reproduced the excitation spectra of $^{11}$N-$^{11}$Be and $^{12}$O-$^{12}$Be mirror nuclei~\cite{Wang2019,Webb}. The wave
functions are constructed in Jacobi coordinates with a complex-energy basis, which exhibit unique three-body features and provide a compelling approach to study the interplay between the single-particle structure and two-neutron correlations. As shown in Fig.~3, the spectra of $^{12}$Be and corresponding $S$ calculated in the framework of GCC are in good agreement with experimental data, including the previously known  $1^-$ state and the currently determined $0^-$ state. 

Furthermore, the width of the $0^-$ state has also been checked using the Wentzel–Kramers–Brillouin (WKB) approximation with single-particle tunnelling through a WS potential  with a reduced radius $r_0 = 1.25$ fm and a diffuseness $a_0=0.65$ fm, plus a centrifugal potential for $\ell =1$. 
Taking into account $S=0.78(23)$, the width of the $0^-$ state was estimated to be 0.012(3)\,MeV, fairly within the experimental upper limit. 
This further confirms the selection of $\ell=1$ transfer in the reaction and the assignment of $J^{\pi} = 0^-$ for the $3.21^{+0.12}_{-0.04}$-MeV state. 

Looking at the excitation energy spectra predicted by the GCC approach, there are three possible candidates ($2^-_1$, $1^-_2$ and $2^-_2$) for the observed $4.44^{+0.15}_{-0.15}$-MeV state. 
However, based on the comparison between the DWBA calculation and the experimental differential cross sections, the $S$ for a pure $1^-$ state is clearly larger than that for a pure $2^-$ state. As a result, the $1^-$ assignment would lead to  an resonance width largely exceeding the experimental limit. Therefore, we adopt a tentative assignment of $2^-$ to the $4.44^{+0.15}_{-0.15}$-MeV resonance.    

\begin{figure}
  \includegraphics[width=1.0\columnwidth]{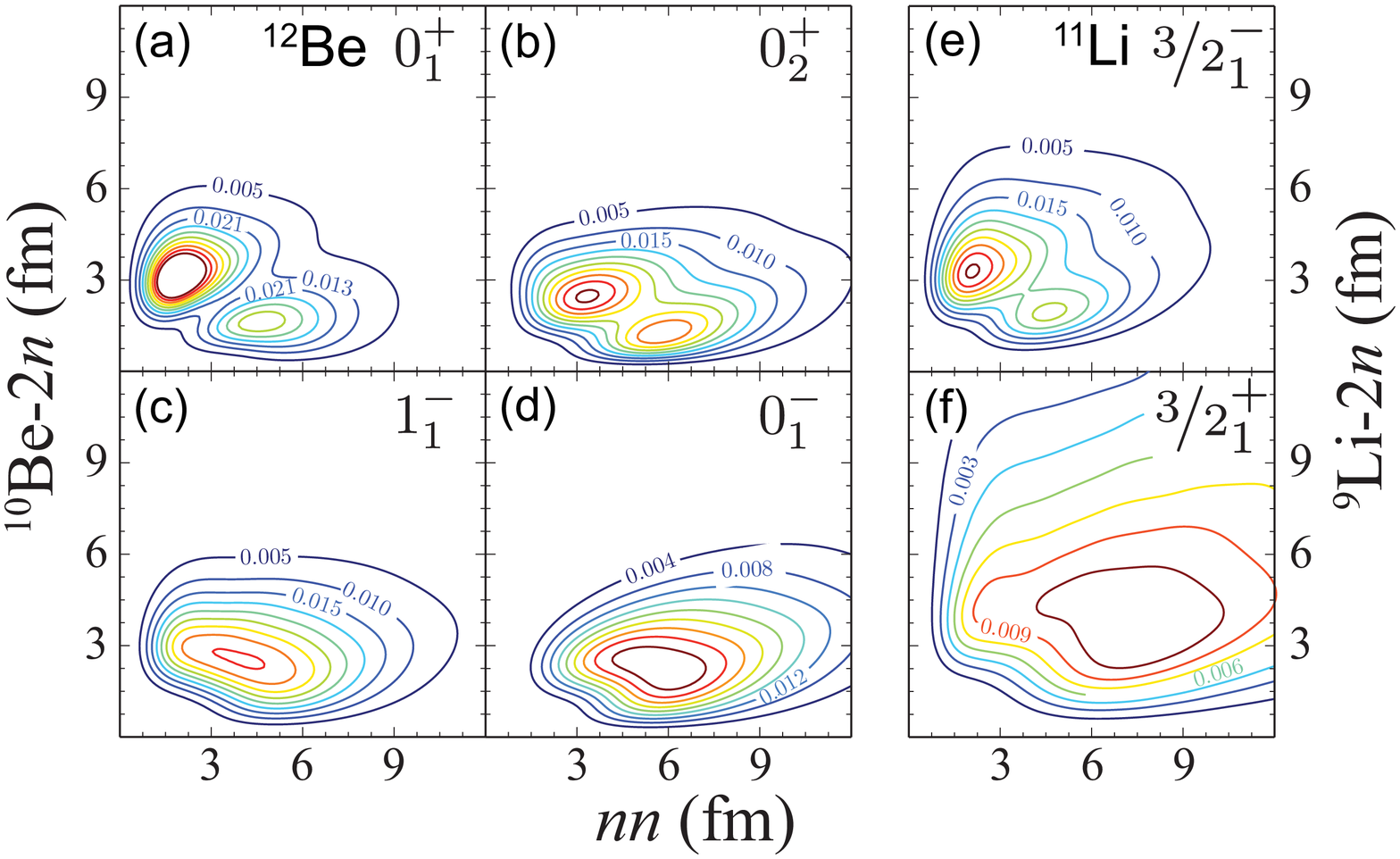}\\
  \caption{Two-nucleon density distributions (in fm$^{-2}$) in Jacobi
coordinates predicted by GCC for the g.s. and low-lying states in $^{12}$Be (a--d) and $^{11}$Li (e--f).}\label{Jacobi_density}
\end{figure}

Since the low-lying $1^-$ and $0^-$ 1p-1h states have been found in $^{14}$C and $^{12}$Be, it is interesting to investigate  similar states in the even less bound isotone $^{11}$Li~\cite{Hagino2005, Wang2019}.  
We show in Fig.~\ref{Jacobi_density} the Jacobi-coordinate density distributions for the g.s. and low-lying states of $^{12}$Be and $^{11}$Li, as calculated in the GCC approach. 
The density distribution of the $^{12}$Be g.s., as shown in Fig.~4(a), exhibits a mixture of  dineutron-like and cigar-like configurations~\cite{Hagino}. 
Its GCC-predicted single-particle configuration is mainly the pairing-featured mixture of $(sd)^2(60\%)+p^2(35\%)$, in line with previous observations~\cite{CHEN2018,chargeex,Pain}.  
The density distribution of the $0_2^+$ state of $^{12}$Be (Fig.~4(b)) still exhibits a two-component distribution, but is more spatially expanded due most likely to the much smaller binding energy and the reduction of the $d$-wave in its configuration ($s^2(16\%)+p^2(57\%)$). This $0_2^+$ state in $^{12}$Be is similar to the two-neutron halo structure of the  $^{11}$Li g.s., shown in Fig.~4(e) with a configuration $s^2(25\%)+p^2(63\%)$. As for the $1^-$ and $0^-$ states in $^{12}$Be (Fig.~4(c) and (d)), the density distributions are diffused into one large component, reflecting the pair-broken 1p-1h configuration. The GCC-predicted main wave function configuration is $(s,p)$ with $78\%$ and $82\%$ probability for the $1^-$ and $0^-$ states, respectively. This large $s$-wave neutron content is reminiscent of the one-neutron halo structure of the $^{11}$Be g.s. ($\sim70\%$ intruder $s$-wave ). We see here the rich effects of the continuum-induced shell melting and cross-shell intrusion, which lead to exotic structures not only in the pair-coupled states, but also in the pair-broken 1p-1h low-lying states at the vicinity of the particle-separation threshold.  
For the isotone $^{11}$Li, according to the GCC calculation, the 1p-1h low-lying resonances are  
the $5/2^+$ ($E_x$ = 0.906\,MeV, $\Gamma=0.258$~MeV) and $3/2^+$ ($E_x = 0.990$\,MeV, $\Gamma=0.277$\,MeV) states~\cite{Webb,Wang2019}, corresponding to $1^-$ and $0^-$ in $^{12}$Be, respectively. 
As an example, we plot the density distribution of the $3/2^+$ state in Fig.~4(f). The overall shape of the distribution is similar to that for $^{12}$Be, but dramatically expanded in space, which may be attributed to the higher relative energy above the particle-separation threshold and the different core property together with the related core-valence interaction~\cite{Webb,Wang2019}. 
The latest finding of the soft dipole resonance might be the predicted 1p-1h states (either $1/2^+$, $3/2^+$ or $5/2^+$)~\cite{Kanungo2015}. Further experimental confirmation of  very exotic 1p-1h resonances in $^{11}$Li would be of special interest. Combining these theoretical and experimental results, we show in Fig.~5 the  binding energy of the 1p-1h states relative to $S_n$, as a function of increasing $Z$ for $N=8$ proton-deficient isotones~\cite{Iwasaki2000,back,14C,Kanungo2015}.  
The nearly linear trend reflects 
the sensitivity of using these states to test  model calculations with  explicit coupling to the continuum. 

\begin{figure}
  \includegraphics[width=1.0\columnwidth]{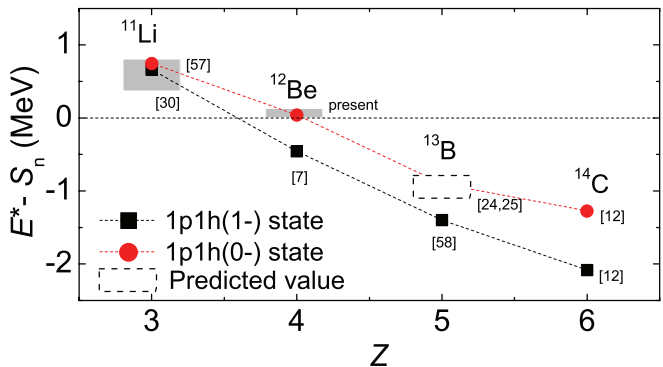}\\
  \caption{The binding energies of the 1$^-$ and 0$^-$ (or $5/2^+$ and $3/2^+$) 1p-1h states as a function of $Z$, for $N = 8$ isotones. The $3/2^+$ state in $^{13}$B has not been found experimentally, and is marked by the theoretical value~\cite{Yuan,WBP}. 
  }\label{protonE}
\end{figure}

It is worth noting that the currently observed $0_1^-$ at $E_x=3.21_{-0.04}^{+0.12}$\,MeV is located very close to the  $S_n$. This means the $0_1^-$ state can only decay to the g.s. of $^{11}$Be through a $p$-wave, which makes the state narrow and relatively stable. 
If the energy of the $0_1^-$ state was higher, the $\ell=0$ decay channel to the $1/2^-_1$ of $^{11}$Be at $E_x = 320$ keV would open, leading to much broader resonance or even a virtual resonance. In fact, it is the $2s_{1/2}$-intruding $^{11}$Be sub-system and the threshold effect of the valence particle, which allows to generate the convenient location for the current narrow $0_1^-$ resonance. 
Recently it has been indicated that the formation of narrow resonances at the close proximity of the corresponding decay threshold is a quite general phenomenon, resulted from the coupling between the internal configuration mixing and the open decay channel mixing~\cite{Okolowicz2020,Ikeda1968}.  These threshold  aligned states may form a relatively stable structure carrying characteristics of the nearby decay channel. The present observation of the narrow near-threshold $0^-$ resonance, together with the extracted sizable $S$, provides a new evidence for this ``alignment" effect, in addition to, for example, the 7.654-MeV Hoyle state in $^{12}$C~\cite{Freer2014,Freer2018}, and the 11.425-MeV state in $^{11}$B~\cite{Ayyad}.

In summary, using the $^{11}$Be\,$(d,p)^{12}$Be reaction, some low-lying unbound states in $^{12}$Be are investigated. A resonance at $3.21^{+0.12}_{-0.04}$\,MeV, just above 1n-separation threshold, was observed for the first time. Based on the DWBA and decay-width analysis, a spin-parity of $0^-$ is assigned to this threshold aligned state, corresponding to a 1p-1h configuration. The successful reproduction of the low-lying states using the GCC approach demonstrates the essential role of the continuum-coupling for weakly-bound and unbound nuclear systems. 
Further studies for more resonant states in $^{12}$Be, $^{11}$Li and other light unstable nuclei are expected in order to promote understanding of the new aspects of  open quantum many-body systems.  

The authors would like to acknowledge the operation staff at RCNP and EN course for providing the beam. The authors from Peking University acknowledge local support. This work is supported by the National Key R$\&$D Program of China (Contract No. 2018YFA0404403), the National Natural Science Foundation of China (Contracts No. 11775004, No. U1867214, No. 11775013, No. 11775316, No. 11875074, No. 11961141003, No. 11835001, No. 11921006, and No. 11975282). This material is based upon work supported by the U.S. Department of Energy, Office of Science, Office of Nuclear Physics under award number DE-SC0013365, Grant No. DE-SC0020451 and Cooperative agreement DE-SC0000661, the State of Michigan and Michigan State University. D.T.T. and C.J. appreciate the support of the Nishimura International Scholarship Foundation and RCNP Visiting Young Scientist Support Program, respectively. 

\bibliography{12Bedp}
\end{document}